# A Spring-Mass-Damper-Based Platooning Logic for Automated Vehicles

**Ardeshir Mirbakhsh**
Ph.D. Student, Transportation Engineering
New Jersey Institute of Technology, University Heights Newark, New Jersey, 07102
Email: am2775@njit.edu

**Joyoung Lee**
Associate Professor, Civil and Environmental Engineering
New Jersey Institute of Technology, University Heights Newark, New Jersey, 07102
Email: jo.y.lee@njit.edu

**Dejan Besenski**
Deputy Director, ITS Resource Center at NJIT
New Jersey Institute of Technology, University Heights Newark, New Jersey, 07102
Email: besenski@njit.edu

Word count: 5,636 words text + 2 tables x 250 words (each) = 6,136 words

**ABSTRACT**

This paper applies a classical physics-based model to control platooning AVs in a commercial traffic simulation software. In Spring-Mass-Damper model, each vehicle is assumed as a mass coupled with its preceding vehicle with a spring and a damper: the spring constant and damper coefficient control spacing and speed adoption between vehicles. Limitations on platooning-oriented communication range and number of vehicles in each platoon are applied to the model to reflect real-world circumstances and avoid overlengthened platoons. The SMD model control both intra-platoon and inter-platoon interactions. Initial evaluation of the model reveals that the SMD model does not cause a negative spacing error between AVs in a harsh deceleration scenario, guaranteeing safety. Besides that, the SMD model produces a smaller positive average spacing error than VISSIM built-in platooning module, which prevents maximum throughput drop. The simulation result for a regular highway section reveals that the proposed platooning algorithm increases the maximum throughput by 10%, 29%, and 63% under 10%, 50%, and full market penetration rate of AVs with 0.5 sec response time. A merging section with different volume combinations on the main section and merging section and different market penetration rates of AVs is also modeled to test inter-platoon spacing policy effectiveness in accommodating merging vehicles. Travel time reductions of 20% and 4% are gained under low MPR of AVs on the mainlane and merging lane accordingly. Meanwhile, a more noticeable travel time reduction is observed in both mainline and merging lanes and under all volume combinations in higher AVs' MPR.

## INTRODUCTION

Due to rapid population growth and the increased number of vehicles, traffic congestion, collisions, and pollution have become leading causes of decreased living standards. Only in 2017 and in the USA, traffic congestion has caused 8.8 billion hours of delay and 3.3 billion gallons of fuel waste, resulting in a total cost of 179 billion USD (1). Traffic congestion has increased between 1 and 3 percent annually from 2008 to 2017 in the USA, and a national congestion cost of 237 billion USD is forecasted for 2025 (1). Data from several studies prove that human errors play a pivotal role in traffic congestion and accidents, and driver error contributes to up to 75% of all roadway crashes worldwide (2). The global rate of road traffic death is 18.2 per 100.000 population. However, there is significant variation across different regions; the USA and Europe hold the lowest regional rates with 15.6 and 9.3 death per 100.000 population, respectively (3).

With advances in communication and sensing technology within the last decade, vehicles are equipped with several driving assistant systems such as Adaptive Cruise Control (ACC), blind-spot monitors, back-up cameras, and lane centering. Some high-end cars produced within the last couple of years can perform most driving tasks without human driver interference, and the imminent widespread appearance of Connected and Autonomous Vehicles (CAVs) on the streets is expected. The most crucial feature of CAVs seems to be the communication capability. The CAVs' communication capability is promising in congestion relief and safety improvement at roadway facilities. Several previous studies have approved that the shorter following gap time between CAVs gained by Cooperative Adaptive Cruise Control (CACC) systems can increase the roadway segment capacity up to 100% (4). It also can improve traffic safety by eliminating human errors, which is one of the leading causes of traffic accidents.

Development of CAV's longitudinal control systems has been a point of interest for more than two decades, and several CACC control systems have been developed within this time. A noticeable portion of studies is focused on the dynamics aspect of CACC, such as vehicle mass, tire friction, vehicle powertrain. These aspects are vital in bringing CACC to fruition yet providing limited insights into the impacts of CACC on the overall traffic network. Alternatively, most of the models developed explicitly for traffic assessment of CACC have missed several critical aspects of platooning such as platoon evolution process, communication range limitations, or interactions between platoons.

In this study, the SMD model is applied to control platooning AVs. The SMD model consists of discrete mass nodes interconnected via a network of springs and dampers. This concept describes how objects reduce their oscillation based on the spring constant, damper coefficient, and mass (5). SMD model is used in several research areas over the last few years, such as analyzing the stability of wind turbines in power and energy system engineering (6), vibration analysis of beam-like structures in mechanical and structure science (7) (8), modeling human organ movements in biomechanics (9), analysis of pulsation in heat pipes in thermal engineering (10), and pedestrian crossing behavior modeling (11). Developing CAV's platooning logics based on Spring-Mass-Damper (SMD) system dates back to the late 1990s, and few researchers have deployed this model for the same purpose since then. SMD model incorporates the most critical dynamic aspect of vehicles, the mass, and covers the platoon evolution process. Other aspects such as communication range, limitations on maximum platoon length, and interactions between platoons are reflected in the model in this study.

The SMD model has been previously used to control platooning AVs (12). The main contribution of this study is developing a more realistic logic by setting a maximum platooning-oriented communication range for the AVs and proposing a real-world-compatible platooning behavior by dividing strings of vehicles into sub-platoons to avoid lengthy platoons and accommodate potential merging vehicles. Meanwhile, the SMD model controls both inter-platoon and intra-platoon interactions. The proposed model is coded into commercial simulation software to facilitate traffic-oriented and potential macroscopic or mesoscopic assessments.

## LITERATURE REVIEW

With the emergence of CAVs and taking advantage of communication capabilities within the last decade, conventional ACC systems have evolved to CACC systems (13). CACC-equipped vehicles can share information with other vehicles and the controller unit; this feature enables the controller to provide safer, smoother, and more natural responses (14). Despite being designed to give the driver more comfort and convenience, CACC can increase traffic throughput and safety by allowing a shorter headway between vehicles and platoon formation (15). Since connectivity between vehicles will be mandatory for the new cars in the USA shortly, CACC systems attract lots of attention from academia and industry (16). A noticeable number of transportation engineering studies are involved with developing CACC longitudinal control system and assessing its impacts on the transportation system. A literature review of developed CACC models and their impact on traffic measures is presented in the following sections.

## CACC Models and their Impact on Traffic Measures

One of the first studies assessing the impact of CACC on traffic flow was conducted by Shaldover et al. (17). In this study, the distribution of time gap between vehicles was collected from human-driven CACC and ACC-equipped vehicles in a field test. According to the test results, in ACC mode, drivers were unlikely to adopt a gap smaller than what they choose in manual driving mode. A simulation platform was built-in AIMSUN to compare ACC and CACC based on the car-following time gap derived from the field test. A simplified version of ACC and CACC car following models were deployed to reduce computational efforts. ACC car-following rules were complying with Nissan cars, and the CACC model was derived from (18), which has two main components: 1) speed control mode, which kept the vehicle's speed close to the speed limit, and 2) the gap control mode, which maintained the desired gap between pair of cars. Simulation results on a single-lane straight freeway under different Market Penetration Rates (MPR) of CACC and ACC revealed that ACC is unlikely to produce a significant change in highway capacity. However, CACC can noticeably increase highway capacity in moderate to high MPRs, as the higher dynamic response gives the driver confidence to adopt a shorter gap setting.

Van Arem et al. (19) studied the impact of CACC on a four-lane highway merging section in the MIXIC microscopic traffic simulation model, which simulates traffic on link-level in a network. The functions for CACC acceleration and distance control were derived from (20). The acceleration was calculated based on distance and speed difference with the ago vehicle. The clearance length was a function of the speed and deceleration capability of the preceding vehicle and the declaration capability of the target vehicle. This study revealed that a high MPR of CACC enhances highway capacity after a lane drop section with a relatively high volume. However, since the communication was limited to longitudinal control, with no control on the length and compactness of CACC platoons, CACC vehicles prevented other vehicles from cutting in. Moreover, the impact of a dedicated lane for CACC vehicles also depends on CACC MPR, a CACC MPR of less than 40% results in more significant speed variance and more shockwaves.

Zonuzy et al. designed a CACC control logic with space-based model parameters to improve CACC platoons merging maneuver (21). The idea was to assign each vehicle on the sub-stream to a pair of vehicles on the mainstream, then adjusting the time gap and velocity of vehicles on the mainstream to accommodate the merging vehicle. The vehicle's control logic was based on nonlinear systems dynamic theory from (22) to guarantee both string stability and single-vehicle stability. The control model inputs were set as pre-designed time gap versus location and velocity versus location profiles. Time gap profiles were designed to ensure vehicles maintain a safe time gap and a minimum deceleration distance to prevent performance drop or safety issues. The numerical simulation results revealed that deploying the target profiles and control logic ensures string stability for platooning CACCs. However, vehicles from the sub-stream need to be positioned at the correct location between the mainstream sub-platoons to ensure that minor errors would not appear at the merging point.

Lee et al. (23) developed a CACC control logic based on a Multi-Objective Optimization Problem (MOOP) approach. Four objective functions were selected to be optimized, 1) Target time headway deviation, to control platoon formation time and make sure the platoon is stabilized under traffic disturbances, 2) Unsafe condition, to make sure each pair of cars attain the minimum required headway to secure safety, 3) Vehicular Jitter, to minimize switch between acceleration and deceleration and avoid a drastic change in any of them and improve comfort, 4) Fuel consumption, to reduce fuel consumption and environmental pollution. A genetic algorithm was deployed to optimize four main objective functions along with additional constraints. The output of optimization was used as control logic for platooning CACC vehicles by controlling their acceleration. Simulation Results of a single lane 14.5 km freeway segment in VISSIM software with three different target time headway values revealed that MOOP CACCs with a longer target time headway yield better performance in time headway deviations. However, a shorter time headway provided more throughput. It was also revealed that MOOP keeps a good balance between all objective functions compared to Single Objective Optimization Problem (SOOP) since the SOOP performed biased with a limited search space for optimization.

## SMD-based Longitudinal Vehicle's Control Logic

In 1998 Eyre et al. (24) used an SMD system with linear characteristics to evaluate platooning AVs' longitudinal string stability properties. They considered two types of interactions: 1) unidirectional, each vehicle being connected only to its predecessor without being affected by its follower, and 2) bidirectional, each vehicle being coupled with both proceeding and following vehicles. Two different spacing policies were examined: 1) constant space policy and 2) speed-dependent spacing policy. Simulation results of autonomous commercial trucks revealed that: 1) the unidirectional controller only achieves stability if it is used with speed-dependent policy, 2) bidirectional controller with the constant spacing policy achieves stability only for a specific range of SMD model hyperparameters, 3) in case of speed-dependent spacing policy, the bidirectional controller showed the most efficient performance if space was only is adjusted with predecessor vehicle.

Contet et al. (25) studied a single platoon of three vehicles controlled by SMD logic, longitudinally and laterally. Two simulation scenarios were developed to analyze trajectory spacing error. 1) while maneuvering to avoid an obstacle and 2) after accommodating a merging vehicle. Simulation results revealed that the platooning vehicles could successfully avoid the obstacle, and no unsafe spacing error is produced in any scenario. A physical experiment was also performed by running soccer robots in a prototype playground, and spacing errors were collected while facing speed constraints or moving in curved paths. The simulation results confirmed the flexibility and adaptability of the model. Neither in the numerical simulation nor the physical experiment, real-world roadway circumstances were not adopted this study.

Munigety et al. (26) performed a sensitivity analysis on the SMD model to its three primary hyperparameters, including mass, spring constant, and damper coefficient. They proved that the spring constant and damper coefficient could represent driver aggressiveness and vehicle stability accordingly. It was also demonstrated that speeding capability reduces as the vehicle's mass increases. By simulating a single lane 300m roadway in MATLAB, speed-flow diagrams for four different vehicle types (motorbike, auto-rickshaw, car, truck) were derived. It was revealed that as the vehicle size increases, the roadway capacity decreases.

Bang et al. (27) developed a strategy for AV platoon evolution by reflecting swarm intelligence theory to the SMD model. The swarm intelligence theory describes animals' clustering behavior, such as bird flocking and fish schooling. It also describes the molecular behavior of materials in different phases, including gas, liquid, and solid. (28). In this study, the spring constant was defined as functions of traffic flow, and the damper coefficient was defined as a function of spring constant, mass, and vehicle response time functions. Simulation results revealed that the spring constant represents the tendency for platoon formation or clustering time, and the damping coefficient represents the stability or oscillation of vehicles. Efficient spring constant and damping coefficient functions for minimizing clustering time in high flow circumstances or providing more freedom for lane changing in low volume circumstances were introduced.

Bang et al. (29) also studied platoon stability at merging sections. The idea was to adjust the spring constant and damping coefficient for the cut-in vehicle and its follower to minimize the cut-in movement's impact. The spring constant and damping coefficient for the cut-in vehicle and its immediate follower were defined as functions of spacing and speed difference between the two vehicles. Consequently, the cut-in vehicle and its follower would maintain a lower spring constant, allowing them to adopt a temporary shorter spacing and higher damper coefficient, which reduces speed disturbance since the vehicle matches the leader's speed more quickly. The simulation results for merging sections showed that the speed and spacing variation and recovery time diminishes significantly with the proposed control.

**Summary**

Although few researchers have considered the SMD model to develop a longitudinal car-following model, this model seems to be a well-fitting logic for this purpose. Unlike transportation engineering-oriented car-following models, the SMD model does not miss the most critical dynamic aspect of vehicles: the mass. Simultaneously, the model is not overinvolved with the vehicles' dynamic aspects and can still be used for traffic assessment purposes. The SMD model can reflect driver aggressiveness and vehicle stability. Some researchers have used the SMD model to develop conventional car-following models (24) (25) (26). Furthermore, few SMD-based CACC models have been proposed, mainly focusing on sensitivity analysis of SMD model hyperparameter and missing the practical aspects of the CACC systems such as maximum communication range and platoon length (27) [28] However, this study intends to develop a more realistic CACC model, by setting limitations on the maximum platooning-oriented communication range and platoon length.

**METHODOLOGY**

This section presents the fundamentals and parameters settings of the SMD model.

**SMD System and Vehicle Platooning**

The SMD model describes how objects maintain desired spacing and speed. It also describes how objects reduce their oscillations based on spring constant, damping coefficient, and object's mass (30). Figure 1 illustrates the SMD system for n objects or vehicles.

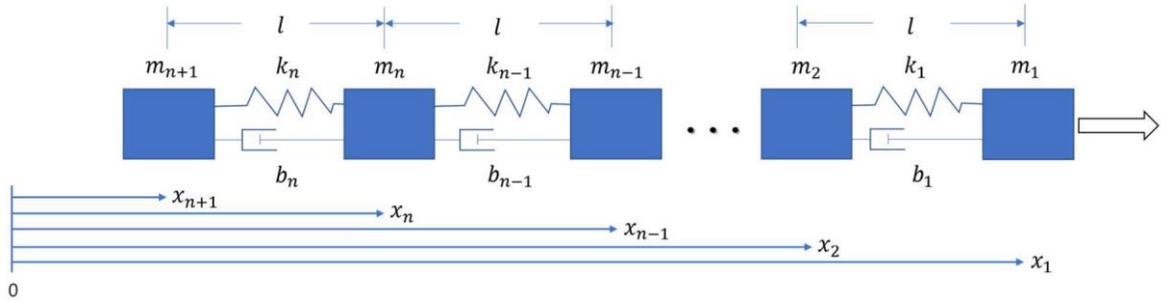

**Figure 1 n-body SMD system (30)**

It is assumed that the friction between vehicles and surfaces is negligible, and each mass is connected only to its predecessor and not affected by its follower. Based on the assumptions, the forces acting on the leading mass and following masses in a string are expressed in Equation (1) and Equation (2) (30).

$$m_1 \ddot{x_1} = c(v_d - \dot{x_1}) \tag{1}$$

$$m_n \ddot{x_n} = k_{n-1}(x_{n-1} - x_n - l) + b_{n-1}(\dot{x}_{n-1} - \dot{x}_n) = k_{n-1}\Delta x_n + b_{n-1}\Delta \dot{x_n} \tag{2}$$

Where:

$m_i$: mass of $i_{th}$ vehicle in the platoon (i = 1 for the lead vehicle)

$x_i, \dot{x}_i, \ddot{x}_i$ : position, speed, and acceleration of $i_{th}$ vehicle

$v_d$ : desired speed of the lead vehicle

$c$: acceleration coefficient

$k_i, b_i$ : spring constant and damping coefficient for the $i_{th}$ vehicle

$l$: Original, unstretched spring length

Parameter $c$ controls the acceleration or deceleration rate of the leading vehicle. Based on the assumption that the maximum acceleration is attained when a standstill vehicle starts moving ($x_1 = 0$), and the maximum deceleration is attained when the lead vehicle wants to stop ($v_d = 0$); constraints (3) and (5) are derived for the $c$ parameter.

$$c \leq \frac{m_1 a_{max}}{v_d} \tag{3}$$

$$c \leq \frac{m_1 d_{max}}{x_1} \tag{4}$$

The variables $a_{max}$ and $d_{max}$ are the maximum acceleration and deceleration capability of vehicles. According to the characteristics of Ford Fusion, 2019 Hybrid, presented in Table 1 (31), $a_{max}$ equals to 3.7 $m/sec^2$, and $d_{max}$ equals to 9.023 $m/sec^2$. Inserting $a_{max}$, $d_{max}$, $m$, and assuming desired speed equal to 120 km/hr.

, $c_{max}$ will be equal to 229.67. To achieve the desired speed in the shortest possible time, the maximum value for $c$ is assumed in this study.

**Table 1 Ford Fusion 2019 Hybrid Specifications**

| Specification | Value |
|---|---|
| Weight | 1676 kg (3695 lb) |
| Length | 4.87 meter (191.7 inches) |
| Acceleration time from zero to 60 mph | 7.3 sec |
| Braking distance 70 mph to zero mph | 178 ft |

The unstretched spring length ($l$) represents the minimum neutral spacing for speed $v$ and is calculated based on Newell's simplified car-following model (32). The $l$ value's function is presented in Equation (5). Considering $l$ as desired spacing for specific speed, the term ($x_{n-1} - x_n - l$) in the model represents deviation from desired spacing.

$$l = s_0 + \tau \times v \tag{5}$$

Where:

$s_0$ : minimum spacing between two vehicles

$\tau$ : response/processing time

$v$ : speed of the target vehicle (the vehicle for which $l$ is being calculated for)

The AVs processing time ($\tau$) ranges from 0.5 to 1 sec (4), while Human-Driven Vehicles (HDV) reaction time ranges from 1.5 to 2 sec (33). The parameter $s_0$ represent bumper to bumper clearance between two standstill vehicles which is assumed equal to 2 meters.

## Characteristics of Spring Constant and Damping Coefficient

The spring constant, $k_i$ represents the spring stiffness. A spring with a larger $k$ is harder to stretch or shrink, but once stretched or shrunk, there is a greater force to recover its original length. Thus, a large $k$ represents a stiff spring with a high body acceleration (high frequency). In contrast, a small $k$ represents a limp spring with low acceleration (low frequency). The spring constant in the SMD model is the coefficient of $\Delta x_n$ and is regarded as an AVs' sensitivity to deviation from its desired spacing with the ago vehicle.

The damping coefficient, $b_i$ represents the degree of resistance that alleviates the spring force. The damping force defines how a following AV approaches its predecessor and adjusts its speed based on the ago vehicle's speed. For example, with a large b, it takes longer for the following vehicle to attain the desired distance, but the speed might be adjusted faster. Meanwhile, with a smaller b, an AV attains the desired distance faster by oscillating in location and speed. The damping coefficient in the SMD model is the coefficient of deviation from desired spacing and can be regarded as an AV's sensitivity to speed deviation from its ago vehicle. The desired spacing is a function of speed, and AV's processing time, derived from Equation (5).

Harmony between the spring constant and damping coefficient is essential for the stability and efficiency of AV platooning. The critical values lead to the shortest time to attain $l$ without collision or oscillations, called critical-damping. Two more possible damping types are 1) under-damping, which is not desirable since vehicles would oscillate back and forth before reaching $l$, and 2) over-damping, which takes longer to reach L, but there is no oscillation. According to [22], conditions of critical-damping and over-damping can be achieved if b $\geq max\left(\frac{m}{\tau}, \sqrt{\frac{k}{m}}\right)$, and the system would be critical-damping if b $= max\left(\frac{m}{\tau}, \sqrt{\frac{k}{m}}\right)$. If we assume that $\Delta x^{\cdot} = 0$, meaning that the force is only coming from the spring, the upper bound for $k$, which provides the greatest spring force, and makes vehicles cluster quickly, is obtained as follows:

$mx^{\cdot\cdot} = ma_{max} = k\Delta x$

$k = \frac{1}{\Delta x} ma_{max}$

Therefore,

$$0 \leq k \leq \frac{1}{\Delta x} ma_{max} \tag{6}$$

## Parameters Setting

A maximum platooning-oriented communication range equal to $4l$ is assumed for AVs. So, as the speed increases, the chance of forming a platoon with a further ago vehicle increases too. Each platoon is divided into several sub-platoons to avoid very long platoons and accommodate potential merging vehicles. According to the sensitivity analysis between platoon length and maximum throughput presented in the proof-of-concept test section, a maximum throughput increment of 3% is gained if the platoon length increases from 4 vehicles to 12 vehicles with a 0.5 sec processing time. Therefore, the maximum platoon length is assumed to be four vehicles to facilitate merging maneuvers. The leading vehicle of each platoon communicates with the last vehicle of the predecessor platoon and, inter-platoon desired spacing is set equal to $3l$.

The maximum value for spring constant and critical value for damper coefficient, presented in Equations (7) and (8), are used in the model to minimize clustering time.

$$k = \frac{1}{\Delta x} ma_{max} \tag{7}$$

$$b = max\left(\frac{m}{\tau}, \sqrt{\frac{k}{m}}\right) \qquad (8)$$

Different vehicle types in the model based on the distance between them and their platooning circumstances are shown in Figure 2.

- Vehicle 1: leading vehicle of a platoon, with having no ago vehicle in the communication range, equation (1) defines this vehicle's acceleration, and it keeps accelerating to maintain the desired speed as long as it does not detect any other vehicle in the communication range.
- Vehicle 2 and 4: following vehicles in a sub-platoon, maintaining the intra-platoon desired spacing ($l$), and the acceleration is derived from Equation (2).
- Vehicle 3: leading vehicle of a sub-platoon, either its sub-platoon or its predecessor sub-platoon, has reached the maximum length. So, this vehicle maintains inter-platoon spacing ($3l$) with its ago vehicle within the communication range, and the acceleration is derived from Equation (2).
- Vehicle 5: having no vehicle in the communication range and maintaining the desired speed based on Equation (1).

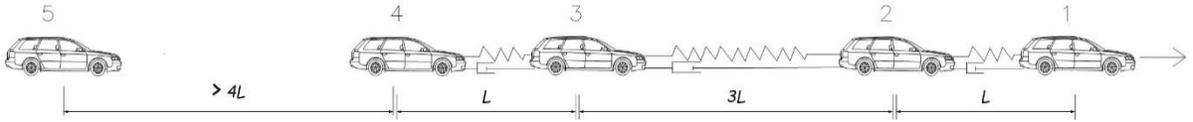

**Figure 2 SMD-based platooning system**

## Microsimulation Testbed Development

The SMD-based AVs' platooning logic is developed in Python and integrated with VISSIM traffic microsimulation software. The VISSIM stochastically produces vehicles and directs them through specified routes. It also controls potential human-driven vehicles based on the Wiedemann 99 car-following model (34), and collects traffic measures such as delay and travel time. The python program collects vehicle data such as location and speed, establishes communication between vehicles, and controls AVs' acceleration based on the SMD model. The Common Object Model (COM) interface allows the python program to access the simulation network's data and objects during simulation. Adjusting editable objects and variables such as traffic volume and individual vehicle's speed is also possible in the COM interface. The process of integrating VISSIM and Python through the COM interface is shown in Figure 3.

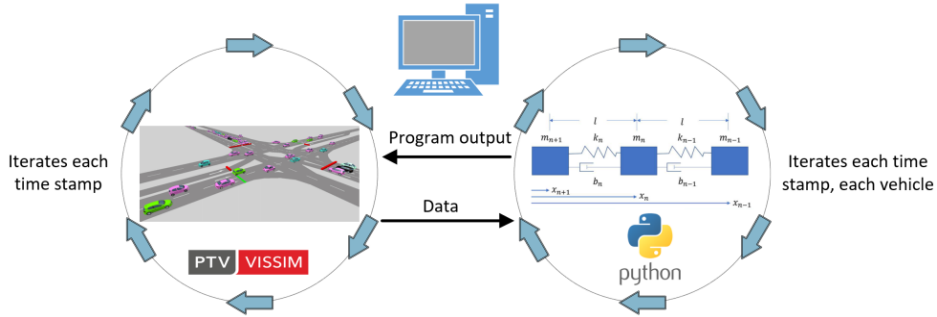

Figure 3 Integration of VISSIM and Python

## PROOF OF CONCEPT TESTS

## Spacing Error

The spacing error between platooning vehicles refers to the difference between actual and desired inter-vehicle spacing. The desired spacing is derived from Equation (5) for the SMD model. The spacing error is expected to be nonzero if the preceding vehicle is accelerating or decelerating (12). A negative spacing error is considered a safety threat, and a positive value will result in a capacity drop. For the initial evaluation of the model, platooning vehicles are exposed to a sudden deceleration scenario, and the average spacing error between AVs is collected. In this scenario, a string of 20 AVs (consisting of 5 platoons of 4 vehicles) is following an HDV which suddenly decelerates with a rate of 5.5 $m/sec^2$ and decrease its speed from 120 km/hr to 30 km/hr. The average spacing error between platooning vehicles is collected over time. VISSIM's built-in A.V. platooning model added to VISSIM since the 2020 version is adopted as a counterpart of the SMD model. However, the details of the VISSIM's built-in platooning model are unknown, considering that modeling another platooning logic in VISSIM is highly time-consuming and requires lots of programming; the VISSIM built-in model is selected as a counterpart of SMD for initial proof of concept test.

The adjustable platooning parameters for the VISSIM built-in model and the SMD model are shown in Table 2; as shown in the table, the parameters are identical between the two models. Except for the maximum platooning-oriented communication range, which is set to 35 meters in the VISSIM built-in model, it is impossible to define this value as a speed function like what we did for the SMD model.

**Table 2 Vissim built-in and SMD model's platooning parameters setting**

| Parameter | VISSIM | SMD |
|---|---|---|
| Maximum number of vehicles in each platoon | 4 vehicles | 4 vehicles |
| Maximum platooning-oriented communication range | 35 meters | $4\,l$ |
| Platoon follow-up gap time | 0.5 sec | 0.5 sec |
| Minimum clearance distance | 2 meters | 2 meters |
| Maximum platooning desire speed | 120 km/hr | 120 km/hr |
| Maximum acceleration | $3.7\ m/sec^2$ | $3.7\ m/sec^2$ |
| Maximume decelaration | $9.023\ m/sec^2$ | $9.023\ m/sec^2$ |

According to the VISSIM2021 manual, vehicles' safety distance is calculated based on Equation (9) (34).

$$PlatoonMinClear + vReference * PlatoonFollowUpGapTm \qquad (9)$$

Where:

*vReference:* The lead vehicle's speed in the platoon

A 200-second simulation is run five times with a 10 step/sec resolution. The simulation's average result is shown in Figure 4.

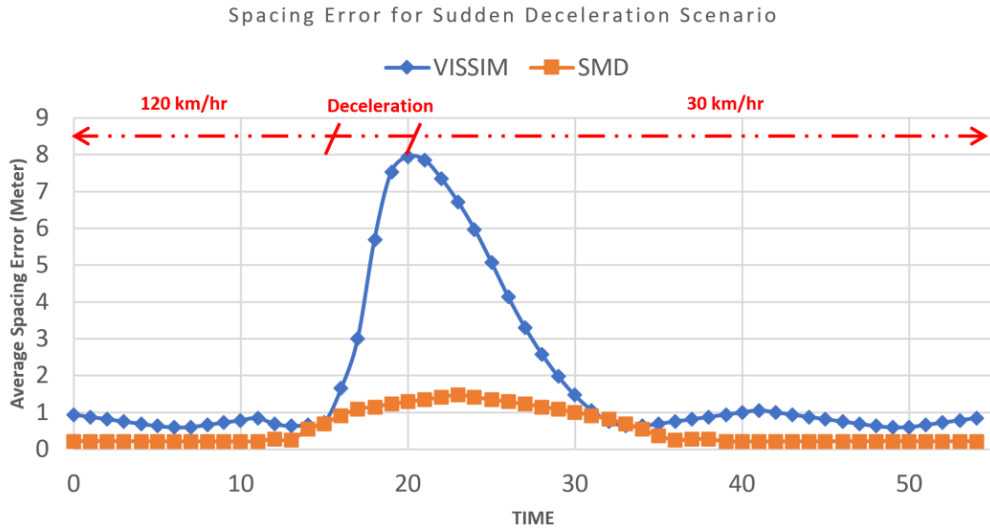

**Figure 4 Spacing error occurrence in harsh deceleration scenario**

According to the results, minor positive errors (less than one meter) are constantly observed for the VISSIM built-in model while following an HDV driver. An increasing error appears as soon as the deceleration phase starts and reaches the climax of 8 meters at the end of the deceleration phase. After 10 seconds of moving with constant speed, the spacing error is reduced to the minimum value. However, no error is observed regularly in the SMD model, and a maximum error of 1.5 meters is observed at the end of the deceleration phase. The advantage of both models is that no negative spacing error is observed, which is equivalent to a safety threat. However, noticeable positive errors are observed in VISSIM built-in model, which results in a maximum capacity drop.

## Maximum Throughput Assessment

A single-lane highway with a length of 4 kilometers is modeled in VISSIM to evaluate the proposed model's impact on the roadway's maximum throughput under different AVs' MPR and processing time ($\tau$). HDVs are controlled by VISSIM's car-following model (Wiedemann 99), and it is assumed that all HDVs are equipped with Vehicle Awareness Device (VAD). VAD is an aftermarket positioning and onboard communication unit proposed by the US Department of Transportation's CVs Initiatives to improve CACC systems' performance in low MPR. A VAD-equipped vehicle broadcasts a Basic Safety Message (BSM), including its location and speed, so its follower AV can use CACC capability (35).

According to Figure 5, assuming a processing time of 0.5 sec for AVs, maximum throughput increments of 4%, 10%, and 17% are gained by the SMD model in low market penetration of AVs, which are 10%, 20%, and 30% accordingly. The maximum throughput increases as AVs' MPR increases, and a maximum throughput increment of 63% is achieved by the SMD model in a full MPR of AVs. The VISSIM built-in platooning module gains slightly more maximum throughput increment than the SMD in higher MPR of AVs. The reason is that the VISSIM built-in module does not follow the inter-platoon spacing policy, resulting in higher occupancy.

As shown in Figure 6, assuming a processing time of 1 sec for the AVs results in a negligible maximum throughput increment in low MPR of AVs, and maximum throughput increments of 23% and 31% are gained by the SMD and Vissim built-in module accordingly.

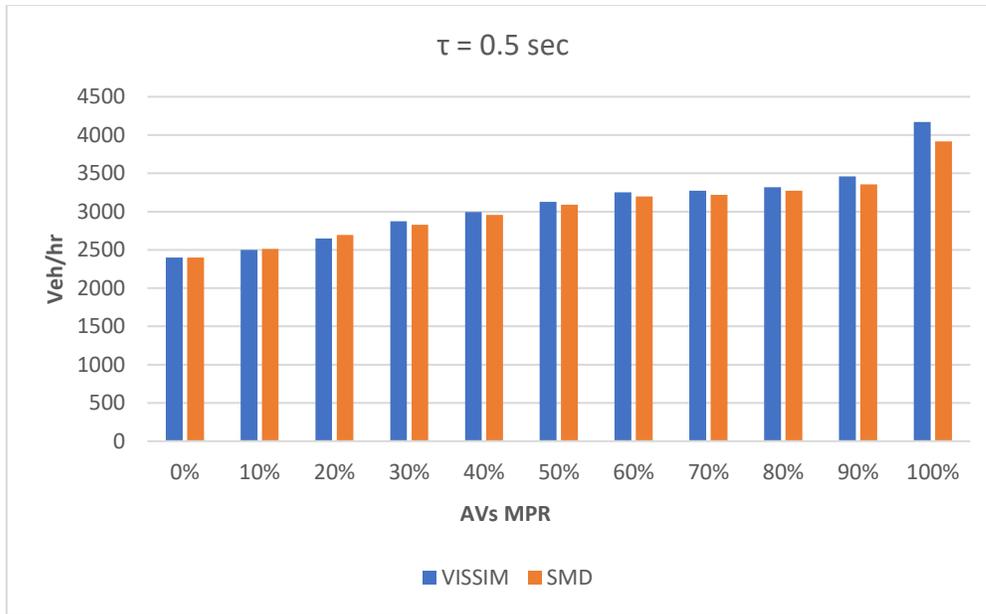

**Figure 5 Comparison of maximum throughput increments between SMD and VISSIM built-in model under different AVs' MPR ($\tau = 0.5\ sec$)**

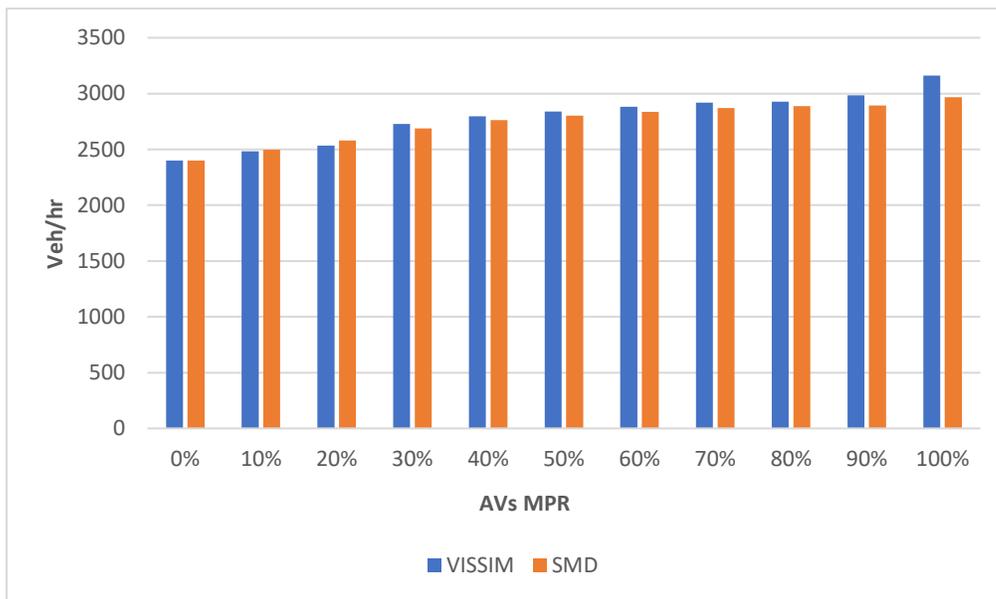

**Figure 6 Comparison of maximum throughput increments between SMD and VISSIM built-in model under different AVs' MPR ($\tau = 1\ sec$)**

As mentioned previously, a sensitivity analysis between the maximum platoon length and maximum throughput is also performed for the SMD model. In this experiment, the maximum platoon length is increased from 4 vehicles to 12 vehicles with an interval of two vehicles. According to the simulation results presented in Figure 7, the maximum throughput slightly increases with platoon length increment. For instance, a maximum throughput increment of 3% is gained if the platoon length increases from 4 vehicles to 12 vehicles with a 0.5 sec processing time.

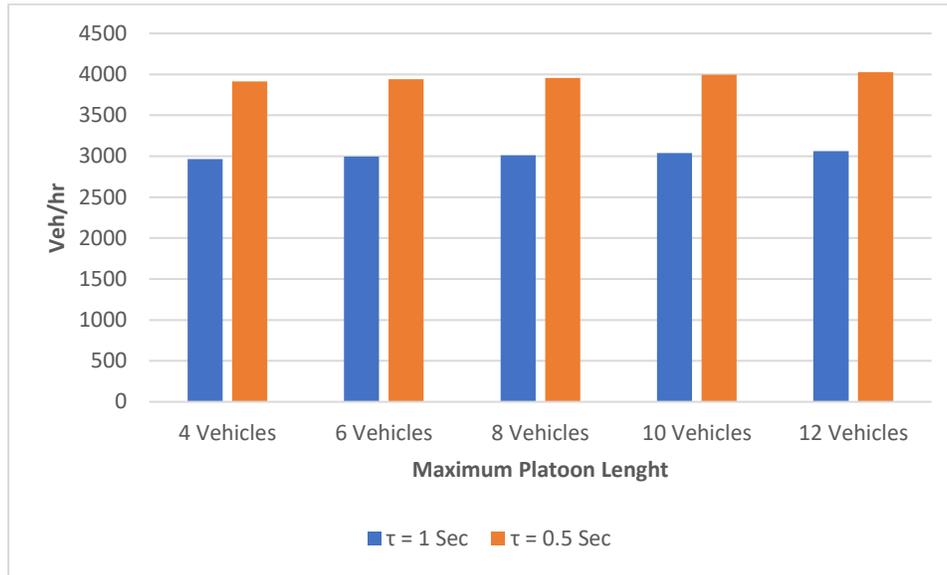

**Figure 7 Sensitivity analysis between platoon length and maximum throughput**

## Travel Time at Merging Section

As shown in Figure 8, a merging section is modeled in VISSIM. In this scenario, AVs on the main lane are in platooning mode, and AVs on the merging lane are controlled by VISSIM built-in car following model (Wiedemann 99) before entering the main lane. Merging vehicles will change lanes as soon as enough space is observed on the main lane. This scenario aims to assess the functionality of the SMD model's inter-platoon spacing policy by comparing its performance with VISSIM built-in model in accommodating merging vehicles. Controlling the merging maneuver is not a point of interest in this study, and lane changing maneuver is performed by VISSIM software. Vehicles from the merging lane start platoon-oriented communication as soon as they enter the main lane to join the existing platoons.

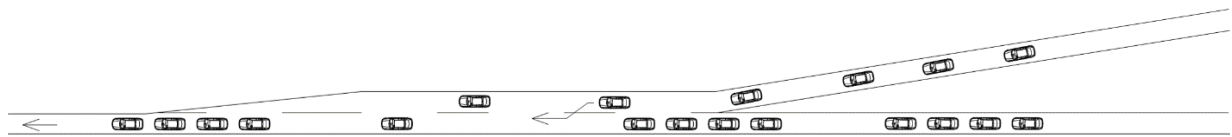

**Figure 8 Merging section scenario**

The maximum speed on the main and merging lanes is 120 km/hr. Traffic volumes on the main lane and merging lane are set to 1800 and 600 veh/hr accordingly, and AVs' processing time is set to 0.5 sec. Simulation time is set to 15 minutes, and the resolution is set to 10 steps/sec. The presented results are the average of running ten simulations with different random seeds. The travel time's data collection length is 280 meters for both merging and main lanes. According to the main-lane travel time results presented in Figure 9, the SMD model gains 24% and 66% reductions assuming 20% and 100% MPR of AVs accordingly. Travel time reductions gained by the VISSIM built-in module on the mainlane are slightly higher than the SMD model, specifically in higher MPR of AVs with 12% more travel time reduction than the SMD model in full MPR of AVs. The reason is that the Vissim platooning module does not follow any inter platoon spacing policy, blocking potential merging vehicles.

Figure 10, which presents merging lane travel time results, deploying the SMD platooning model decreases travel time by 3% in AVs' MPR of 20%, and a maximum travel time reduction of 36% is gained in full MPR of AVs. However, the Vissim built-in platooning logic has caused a noticeable travel time increment on the merging lane, specifically in higher MPR of AVs. For instance, 10% and 37% travel time increment compared to the none platooning circumstances is observed in 40% and full MPR of AVs accordingly.

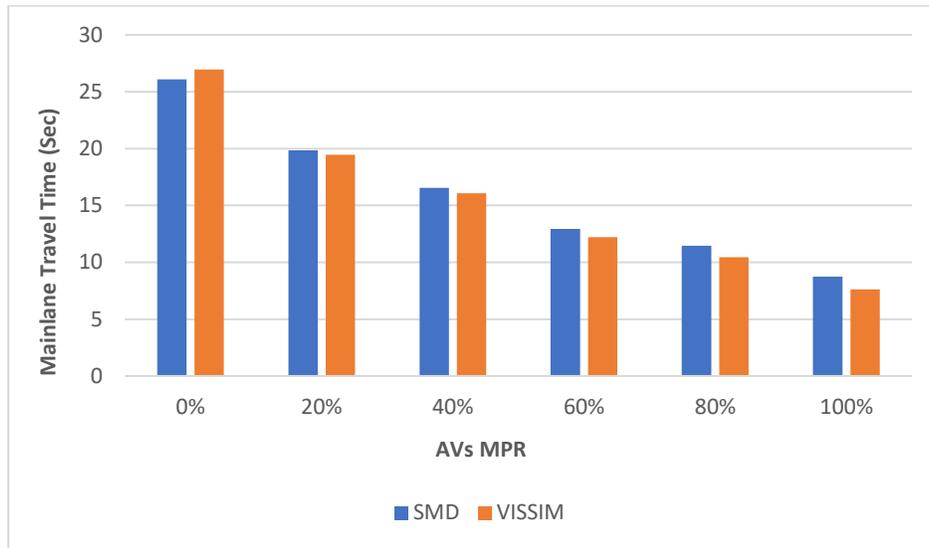

**Figure 9 Mainlane travel time comparison between SMD and VISSIM built-in module**

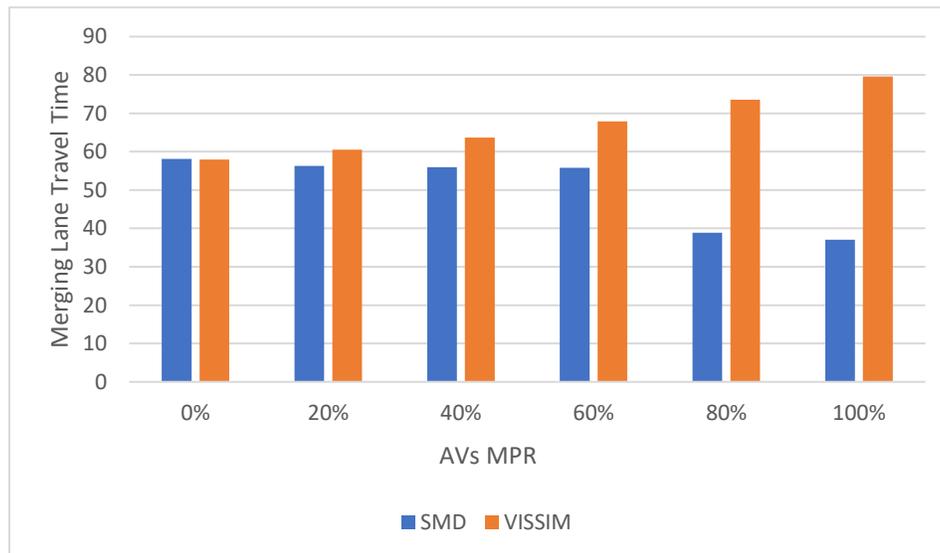

**Figure 10 Merging lane travel time comparison between SMD and VISSIM built-in module**

## CONCLUSIONS AND FUTURE DIRECTIONS

This paper presented an SMD-based platooning logic for AVs. Limitations are set on AV's maximum platooning-oriented communication range to reflect real-world circumstances into the model and save communication and processing resources. Moreover, to avoid overlengthened platoons, long strings of AVs are divided into sub-platoons, and the SMD model controls both inter-platoon and intra-platoon interactions. The model is coded into a commercial traffic simulation software, facilitating transportation-oriented and potential macroscopic or mesoscopic assessments.

As an initial evaluation, the developed model was compared with VISSIM 2020 built-in platooning module. Simulation results revealed that the SMD model produces 75% less positive average spacing error than the VISSIM 2020 built-in platooning module in a harsh declaration scenario, which guarantees more throughput in unstable driving circumstances. Moreover, no safety threat is detected in any of the models' extreme deceleration maneuvers since no negative spacing error is observed.

Modeling a single lane highway section revealed that the SMD-based platooning AVs gain a maximum throughput increment of 63% and 27% in the full and 50% MPR of AVs with a response time of 0.5 $sec$, and a maximum throughput increment of 23% and 17% was achieved in the full and 50% MPR of AVs with a response time 1.0 $sec$. Assuming a lower MPR of AVs, maximum throughput increments of 4% and 10% were gained in 10% and 20% MPR of AVs, with 0.5 sec processing time. A highway merging section was modeled with platooning AVs on the main lane, and merging vehicles moving to the main lane as soon as enough space was provided for them. In this scenario, travel time reductions of 50% and 67% on the main lane were achieved in 60% and 100% MPR of AVs, and noticeable travel time reduction on the merging lane was observed if AVs' MPR is 80% and higher, and the merging volume is high. Assuming AVs' MPR of 20%, 3%, and 20%, travel time reductions were gained on the merging lane and mainlane accordingly. Considering Vissim built-in platooning logic as a counterpart for the proposed model, the effectiveness of the proposed model's inter-platoon spacing policy in accommodating merging vehicles was confirmed.

Potential future studies could include adding a lane change logic to the model, modeling multilane highway sections, and finding the optimum platoon length for different traffic circumstances. Modeling varied AVs types with different characteristics such as mass, response time, and acceleration or deceleration rates can also be considered.